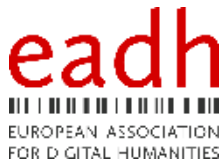

# Visual Reconstruction as Memory Negotiation: An Iterative Generative AI-Mediated Framework for Oral History

Yigeng Zhang [1]

[1]Independent Scholar

## 1 Introduction

Oral history and community memory are core resources for historical inquiry. Across settings with uneven archival and media infrastructures, elders' oral accounts often remain the most accessible testimony for recent social change. In contexts with fewer surviving visual records, a condition referred to here as visual archive scarcity, audiences depend on narrated testimony to imagine past environments. Even vivid storytelling can leave spatial relations and material details under specified. This reflects differences in how verbal and visual representations organize information, with narrative language foregrounding meaning and affect while images convey spatial configuration and fine-grained cues. These modes are complementary, yet multiple narrators' recollections can diverge, and verbal exchange alone may not make those differences easy to compare, document, or negotiate.

## 2 Theoretical Framing

This framework starts from a shared premise in oral history and memory studies: recollection is meaningful not because it reproduces facts, but because it organizes experience through narrative interpretation and social framing (Portelli 2010; Halbwachs 2020; Assmann 2011). Accordingly, the goal is not to *verify* memory via images, but to *externalize* it in a form that can be discussed, revised, and compared. At the same time, work in digital humanities warns that visual and computational forms can naturalize interpretation as evidence and thereby exercise *visual authority* (Drucker 2011). These commitments motivate a methodological choice: generative AI is used as a speculative, revisable medium for negotiation, with transparent labeling and archived iterations, rather than a device for producing definitive reconstructions. Related practice-based work, such as the Synthetic Memories project, indicates the practical use of participatory, interview-grounded, iterative image-making in memory-oriented settings, while also highlighting the need to frame evidentiary claims cautiously (Bañuelos Capistrán, Zavala Scherer, and Lugo Rodríguez 2025).

## 3 Method: An Iterative Framework

The proposed Iterative AI-Assisted Framework for Visual Reconstruction and Memory Negotiation unfolds in five stages. Figure 1 shows an illustrative vignette of this framework.

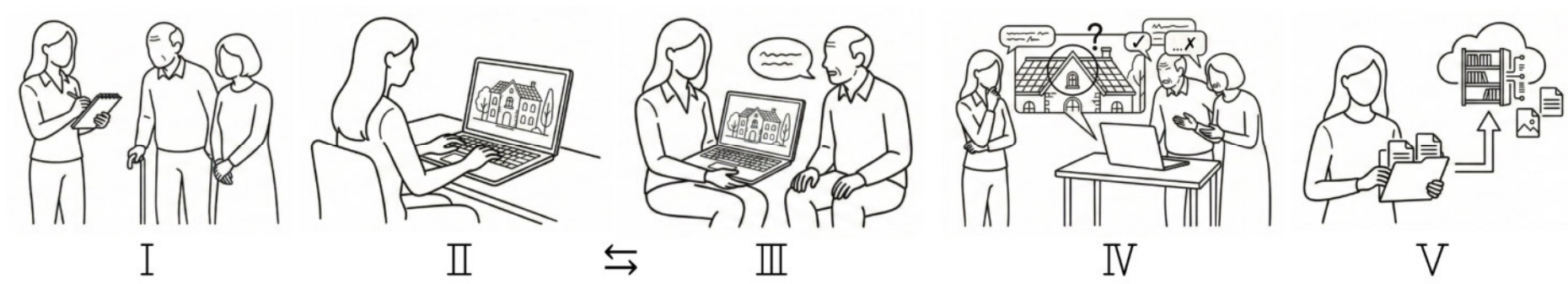

Figure 1: Vignette of iterative AI mediated visual reconstruction as memory negotiation. An interviewer and two elders externalize memories of a no longer extant building. Narrative elicitation informs a provisional draft, revised with participant feedback, compared across narrators, and archived with intermediate versions and notes of agreement, uncertainty, and disagreement. The goal is not evidentiary reconstruction but making interpretive differences discussable and preservable as part of the record.

*Step 1: Narrative Elicitation.* One or more narrators offer a situated account of a scene through guided conversation. The aim is to articulate salient spatial, material, and affective cues, establishing a shared point of reference for subsequent interpretation.

*Step 2: Generative Visual Prototyping.* Text-to-image generation is used to produce an initial draft of the narrated scene. Rather than functioning as reconstruction, the image serves as a provisional *visual probe* that externalizes aspects of the narration for collective inspection.

*Step 3: Participant-Led Iterative Revision.* Narrators respond to the draft by indicating mismatches, omissions, and points of emphasis, and the image is revised through rounds. Intermediate versions are retained to document how recollection is articulated and reshaped through the process.

*Step 4: Multi-Narrator Comparison and Negotiation.* Where multiple witnesses are available, iterated drafts are brought into conversation. Convergences and divergences are discussed explicitly, with differences treated not as noise to remove but as meaningful traces of memory's situatedness and the social work of remembering.

*Step 5: Negotiated Reconstruction Archive.* The resulting images, along with intermediate iterations and a concise record of agreements, uncertainties, and disagreements, is curated as a multimodal archive. The archive foregrounds negotiation and interpretive plurality rather than a definitive visual outcome.

## 4 Discussion and Contribution

This framework is an experimental method that foregrounds mediation and authority. Generative images can appear evidential, with anchoring effects and model confabulations shaping what narrators and audiences *see* as plausible. Therefore, images are treated as provisional artifacts for critique, revision, and comparison rather than claims about what the past *really* looked like. Retaining intermediate iterations and recording moments of disagreement keeps uncertainty visible and makes the interpretive process legible. Future work will require systematic evaluation of participant experience, ethical safeguards, and longer-term interpretive effects.

By reframing visual reconstruction as negotiation, the paper offers a methodological model for integrating generative media into oral history without turning representation into fact. The contribution is a structured way to externalize, discuss, and document memory in plural form, capturing both convergence and divergence as part of the record. Where visual documentation is uneven or absent, this approach supports a reflexive mode of historical engagement that remains attentive to the politics of images and the social life of remembering.